\documentclass[english,prl,a4paper,amsmath,amssymb,showkeys,superscriptaddress,twocolumn,showpacs,floatfix]{revtex4-1}
\usepackage{graphicx}
\usepackage{dcolumn}
\usepackage[colorlinks=true,linkcolor=blue ,citecolor=blue,urlcolor=blue]{hyperref}
\usepackage{multirow}
\usepackage[usenames,dvipsnames]{xcolor}
\usepackage{soul}
\usepackage{braket}
\usepackage{bm}
\usepackage{nicefrac}
\usepackage{siunitx}
\usepackage{color,transparent}
\usepackage[utf8]{inputenc}
\usepackage{upgreek}
\usepackage[export]{adjustbox}
\usepackage{pict2e}

\newcommand{\VEC}[1]{\mathbf{#1}}

\DeclareSIUnit\ML{ML}
\DeclareSIUnit\MLs{MLs}
\DeclareSIUnit\meVA{meV\angstrom^2}

\begin{document}

\title{Paramagnetic fluctuations of the magnetocaloric compound MnFe$_4$Si$_3$} 

\author{N. Biniskos}
\email{nikolaos.biniskos@matfyz.cuni.cz}
\affiliation{Charles University, Faculty of Mathematics and Physics, Department of Condensed Matter Physics, Ke Karlovu 5, 121 16, Praha, Czech Republic}
\affiliation{Forschungszentrum J\"ulich GmbH, J\"ulich Centre for Neutron Science at MLZ, Lichtenbergstr. 1, D-85748 Garching, Germany}
\author{K. Schmalzl}
\affiliation{Forschungszentrum J\"ulich GmbH, J\"ulich Centre for Neutron Science at ILL, 71 Avenue des Martyrs, F-38000 Grenoble, France}
\author{J. Persson}
\affiliation{Forschungszentrum J\"ulich GmbH, J\"ulich Centre for Neutron Science (JCNS-2), JARA-FIT, D-52425 J\"ulich, Germany}
\author{S. Raymond}
\email{raymond@ill.fr}
\affiliation{Universit\'e Grenoble Alpes, CEA, IRIG, MEM, MDN, F-38000 Grenoble, France}

\date{\today}

\begin{abstract}
Inelastic neutron scattering technique is employed to investigate the paramagnetic spin dynamics in a single crystalline sample of the magnetocaloric compound MnFe$_4$Si$_3$.
In the investigated temperature range, 1.033$\times T_{C}$ to 1.5$\times T_{C}$, where $T_C$ is the Curie temperature, the spin fluctuations are well described by the ferromagnetic Heisenberg model predictions.
Apart from the Heisenberg exchange, additional pseudo-dipolar interactions manifest through a finite long-wavelength relaxation rate that vanishes at the transition temperature ($T_C = 305$\,K).
Based on the characteristic extend of spin fluctuations in wave-vector and energy space we determine that the nature of magnetism in MnFe$_4$Si$_3$ is localized above room temperature.
This contrasts with the most celebrated Mn and Fe based magnetocaloric materials that are considered as itinerant magnets.
The field dependence of the paramagnetic spectra shows a strong suppression of the quasi-elastic excitations, while a field induced spin-wave mode appears at finite energy transfers for a magnetic field of 2\,T. 
This modification of the spectra suggests a decrease of magnetic entropy with applied magnetic field that finds echo in the magnetocaloric properties of the system.
\end{abstract}

\maketitle

\section{Introduction}

Magnetism in solids originates from the atomic magnetic moments, which are a consequence of the spin and angular momenta of the electrons.
A magnetically ordered phase is described by the order parameter $M$ (e.g. magnetization for a ferromagnet) that vanishes in the paramagnetic (PM) state. 
At finite temperature, thermal spin fluctuations mostly account for the reduction of $M$. 
Such fluctuations are enhanced near the transition temperature and persist in the PM state~\cite{Moriya1985}. 
They manifest as a continuum of excitations in the wave-vector - energy ($\bf{q}$, $E$) space.

For several cubic ferromagnets (FM) the spin fluctuations around and above the Curie temperature ($T_C$) have been extensively investigated by inelastic neutron scattering (INS) measurements~\cite{Wicksted1984,Steinsvoll1984,Ishikawa1985,Kohgi1986,Boni1987,Shirane1987,Hiraka1997,Semadeni2000,Kindervater2017}.
These reports retrieved properties about the temperature dependence of the $q$-dependent relaxation rate $\Gamma_{q}$ and the inverse spin correlation length $\kappa$.
At $T_C$, $\Gamma_{q}$ follows the expression $\Gamma_{q} \propto Aq^{z}$, $z$ being the dynamical exponent, while for $T  > T_C$, $\kappa(T)=\kappa_{0}(T/T_{C}-1)^{\nu}$, where $\kappa_{0}$ refers to the inverse spin correlation length at $T=0$\,K and $\nu$ is the usual critical exponent describing the divergence of the correlation length $\xi \sim \kappa^{-1}$ at $T_C$. 
The interest of such studies lies in the fact that when $A$ and $\kappa_{0}$ are compared to $T_{C}$ and the interatomic distance $d^*$, respectively, further insights about the nature of the magnetism of a material can be obtained.
For localized magnetic systems it is expected $A/T_{C} \approx 1$ and $d^*\kappa_{0} \approx 1$, while for itinerant magnets $A/T_{C} >> 1$ and $d^*\kappa_{0} << 1$~\cite{Moriya1985,Endoh2006}.
These differences are due to the fact that the length scales are related to typical interatomic distances in localized magnetic systems and their energy scales are mainly controlled by the exchange interactions, to which $T_C$ scales in a mean-field approach. In contrast other parameters, characteristics of the electronic band structure and the Fermi surface, need to be considered for itinerant systems.
It is also worth mentioning that investigations of the spin fluctuations with INS have focused mainly on cubic FM, while similar studies for FM materials with lower crystal symmetries are scarce~\cite{Raymond2010,Stock2011,Haslbeck2019}.

Hence, the extension of the spin fluctuations in momentum/energy space is a way to characterize the nature of magnetism in solids. 
Apart from being of fundamental interest for understanding the microscopic magnetism of a compound, the fluctuations can play an important role in macroscopic phenomena associated to a given magnetic functionality of a material.
The appearance of significant spin fluctuations can be associated with large changes in the magnetic entropy of a system.
Based on the Maxwell relations it is known that any temperature dependence of the magnetization is connected to an entropy change when changing an external magnetic field.
This effect is exerted for achieving sub-Kelvin temperatures through the adiabatic demagnetization of paramagnetic salts, and for magnetic refrigeration applications in daily life via the giant magnetocaloric effect (MCE) that is observed near room temperature magnetic phase transitions~\cite{Bruck2024}.

Despite the numerous research articles that quantify the MCE in several magnetic materials using macroscopic measurements~\cite{Franco2012,Gschneidner2000,Bruck2024}, such as heat capacity and magnetization, studies of the spin fluctuations and their relation to the MCE have only been discussed in a handful of systems with neutron spectroscopic studies, namely in MnFe$_{4}$Si$_{3}$~\cite{Biniskos2017}, Mn$_{5}$Si$_{3}$~\cite{BiniskosN2018}, (Mn,Fe)$_{2}$(P,Si)~\cite{Miao_2016}, LaFe$_{13-x}$Si$_{x}$~\cite{Faske_2020,Zhang_2021,Morrison2024}, and HoF$_{3}$~\cite{Dixey_2023}.
To this aim, we employ polarized INS measurements and unpolarized INS under magnetic field to probe the magnetization dynamics above $T_{C}$ of the hexagonal metallic FM MnFe$_4$Si$_3$ and we propose links between the underlying temperature and magnetic field effects with the MCE.

In MnFe$_{4}$Si$_{3}$, a second order phase transition from the PM state to the FM phase occurs near room temperature ($T_C = 305$\,K)~\cite{Biniskos2017,Bouachraoui2019,Singh2020}.
The material crystallizes in the hexagonal space group $P6_{3}/mcm$~\cite{Binczycka1973,comment}.
Within the structure, the Wyckoff position (WP) 4$d$ is occupied by Fe atoms surrounded by six Si atoms, while the WP 6$g$ has a mixed occupancy of Fe--Mn (Fe occupancy $\approx67$\%, Mn occupancy $\approx33$\%).
The magnetic moments on the WP 6$g$ lie in the $ab$-plane of the hexagonal symmetry with a magnitude of 1.5(2)\,$\mu_{B}$, while no significant magnetic moment could be determined on the WP 4$d$~\cite{Hering2015}. 

\section{Experimental details}

A MnFe$_{4}$Si$_{3}$ single crystal (the same used in previous studies~\cite{Biniskos2017,Raymond_2019}) with mass of about 7\,g and grown by the Czochralski method was oriented in the $(a^*,c)$ scattering plane of the hexagonal symmetry.
In this article we use the hexagonal coordinate system and the scattering vector $\VEC Q$ is expressed in $\VEC Q = (Q_{h}, Q_{k}, Q_{l})$ given in reciprocal lattice units (r.l.u.). 
The wave-vector $\VEC q$ is related to the momentum transfer through $\hbar\VEC Q = \hbar\VEC G + \hbar\VEC q$, where $\VEC G$ is a Brillouin zone center and $\VEC G = (h, k, l)$.

INS measurements were carried out at the Institut Laue-Langevin (ILL), in Grenoble, France.
Polarized INS data were obtained on the CRG-J\"ulich and CRG-CEA Grenoble cold and thermal neutron three-axis spectrometers (TASs) IN12~\cite{schmalzl2016} and IN22, respectively.
Both TASs were set up in a W configuration and inelastic scans were performed with constant $k_f$ (1.8\,$\text{\AA}^{-1}$ for IN12 and 2.662\,$\text{\AA}^{-1}$ for IN22), where $\VEC{k}_f$ is the wave-vector of the scattered neutron beam. 

On IN12, the incident neutron beam spin state was prepared using a transmission polarizing cavity located after the velocity selector and the initial wave-vector was selected by a double focusing pyrolytic graphite [PG(002)] monochromator. 
For IN22 the spin of the incident neutrons was polarized with a verticallly focusing Heusler [Cu$_2$MnAl(111)] monochromator. 
For both TASs the neutron polarization and $k_f$ were analyzed using a horizontally focusing Heusler analyzer.
All along the neutron path guide fields were installed to maintain the polarization of the beam. 
High-order harmonics were suppressed using a velocity selector before the monochromator for IN12 and a PG filter in the scattered neutron beam for IN22.
A flipping ratio of about 22 and 14 has been measured on a graphite sample on IN12 and IN22, respectively.
In order to perform longitudinal polarization analysis (LPA), we used a Helmholtz coils setup to control the direction of the polarization on the sample.
This method is restricted to temperatures above $T_C$ since the beam gets depolarized when entering the ferromagnetic phase (see the inset of Fig.~5 in Ref.~\onlinecite{Biniskos2017} for the temperature dependence of the beam polarization).
For investigating the spin dynamics above $T_C$ the MnFe$_{4}$Si$_{3}$ single crystal was placed inside a cryofurnace. 

As a general rule, neutron scattering is only sensitive to magnetic excitations perpendicular to $\VEC Q$~\cite{Squires_2012} and by employing LPA it is possible to separate magnetic fluctuations polarized along different directions in spin space~\cite{chatterji_neutron_2006}. 
We use the standard $(x, y, z)$ frame where the $x$-axis is parallel to $\VEC Q$, the $z$-axis is vertical, and the $y$-axis is perpendicular to $\VEC Q$ and $z$.
The corresponding measurement channels where we collected data are canonically labeled NSF$_{xx}$, NSF$_{yy}$, and NSF$_{zz}$, where NSF stands for ``Non Spin-Flip''. 
The neutron scattering double differential cross-sections for the three NSF channels are~\cite{chatterji_neutron_2006}:
\begin{eqnarray}
\text{NSF}_{xx}=\left(\frac{d^2\sigma}{d\Omega d\text{E}}\right)_{\text{NSF}}^{x} \propto \text{BG}_{\text{NSF}} + \langle{N}\rangle \\
\text{NSF}_{yy}=\left(\frac{d^2\sigma}{d\Omega d\text{E}}\right)_{\text{NSF}}^{y} \propto \text{BG}_{\text{NSF}} + \langle{N}\rangle + \langle{\delta}{M}_{y}\rangle\\
\text{NSF}_{zz}=\left(\frac{d^2\sigma}{d\Omega d\text{E}}\right)_{\text{NSF}}^{z} \propto \text{BG}_{\text{NSF}} + \langle{N}\rangle + \langle{\delta}{M}_{z}\rangle ,
\end{eqnarray}
where $\text{BG}_{\text{NSF}}$ is the background, $\langle{N}\rangle$ is the nuclear scattering, and $\langle{\delta}{M}_{i}\rangle$  (with $i=y,z$) the magnetic fluctuations. 

Considering that the scattering plane in our case is $(a^*,c)$ then for $\VEC Q$ parallel to the $(h00)$ direction the scattering cross sections are:
\begin{eqnarray}
\text{NSF}_{xx} \propto \text{BG}_{\text{NSF}} + \langle{N}\rangle \\
\text{NSF}_{yy} \propto \text{BG}_{\text{NSF}} + \langle{N}\rangle + \langle{\delta}{M}_{c}\rangle\\
\text{NSF}_{zz} \propto \text{BG}_{\text{NSF}} + \langle{N}\rangle + \langle{\delta}{M}_{b}\rangle
\end{eqnarray}
and for $\VEC Q$ parallel to the $(00l)$ direction the scattering cross sections are:
\begin{eqnarray}
\text{NSF}_{xx} \propto \text{BG}_{\text{NSF}} + \langle{N}\rangle \\
\text{NSF}_{yy} \propto \text{BG}_{\text{NSF}} + \langle{N}\rangle + \langle{\delta}{M}_{a^*}\rangle\\
\text{NSF}_{zz} \propto \text{BG}_{\text{NSF}} + \langle{N}\rangle + \langle{\delta}{M}_{b}\rangle.
\end{eqnarray}
Therefore, one can extract the magnetic fluctuations polarized along different directions by canonical subtraction of intensities in different NSF channels.

In addition, unpolarized INS data under magnetic field were obtained at IN12 with $k_f=1.8$\,$\text{\AA}^{-1}$.
The PG(002) monochromator was vertically focused and an horizontally focused PG(002) analyzer was used, and 40'-open-open collimations were installed.
The MnFe$_{4}$Si$_{3}$ single crystal was placed inside a 2.5\,T vertical field magnet. 
The magnetic field was applied parallel to the $b$-axis of the hexagonal system of the sample (perpendicular to the $(a^*,c)$ plane).

\section{Experimental results}

\subsection{Temperature dependence of paramagnetic scattering without magnetic field}

To investigate the spin dynamics of MnFe$_{4}$Si$_{3}$ above the Curie temperature, spectra were collected at different temperatures, namely 315\,K, 336\,K, 367\,K, 396\,K, 427\,K, and 457\,K, which is about 1.033$\times T_{C}$, 1.1$\times T_{C}$, 1.2$\times T_{C}$, 1.3$\times T_{C}$, 1.4$\times T_{C}$, and 1.5$\times T_{C}$, respectively. 
Constant $\VEC Q$-scans were carried out at energy transfers $-4\leq E \leq 4$\,meV around the Brillouin zone centers $\VEC G = (2, 0, 0)$ and $\VEC G = (0, 0, 2)$ and data were obtained along the $(h00)$ and $(00l)$ directions.

\begin{figure}[h]
    \includegraphics[width=\columnwidth]{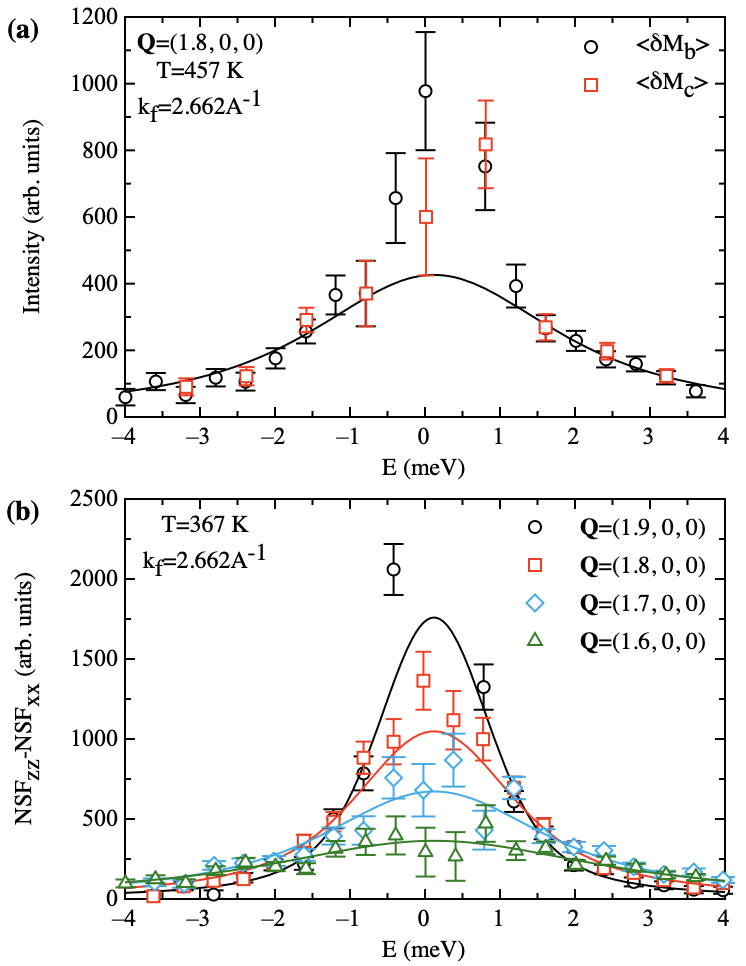}
    \caption{\label{fig:rawdata}
    (a) $\langle{\delta}{M}_{b}\rangle$ and $\langle{\delta}{M}_{c}\rangle$ spin fluctuations spectra of MnFe$_4$Si$_3$ from constant $\VEC Q$-scans at $\VEC Q = (1.8, 0, 0)$ at $T = 457$\,K.
    (b) $\langle{\delta}{M}_{b}\rangle$ spin fluctuations spectra obtained at different $Q_h$ positions along the $(h00)$ direction at $T = 367$\,K.
    Solid lines represent fits as explained in the text.
    }
\end{figure}

\begin{figure}[h]
    \includegraphics[width=\columnwidth]{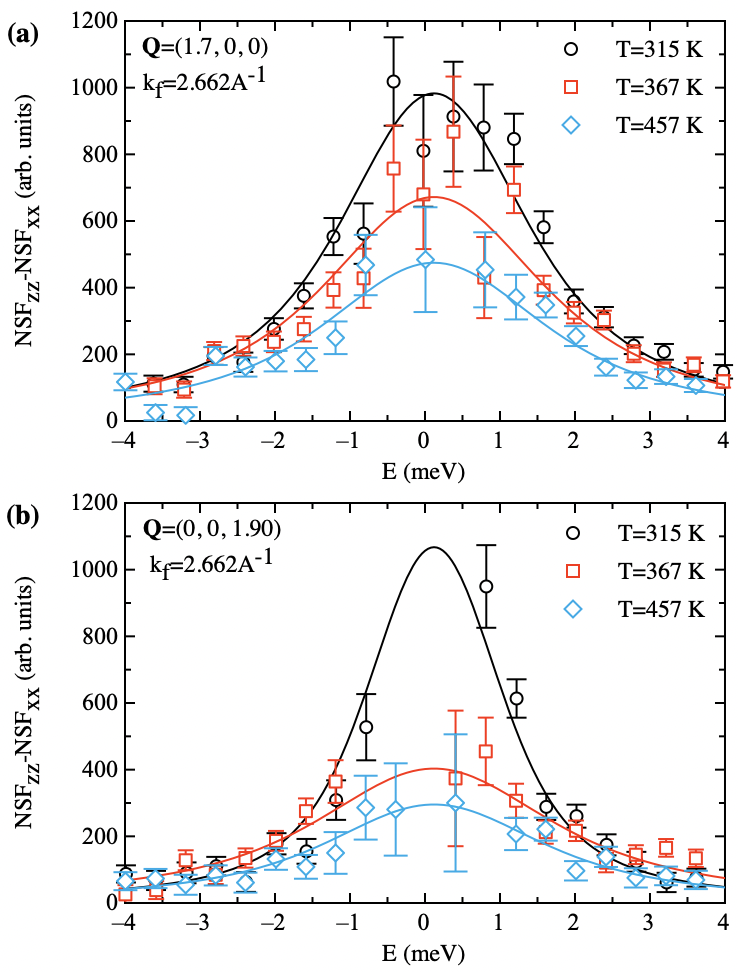}
    \caption{\label{fig:rawdata2}
    $\langle{\delta}{M}_{b}\rangle$ spin fluctuations spectra of MnFe$_4$Si$_3$ obtained at different temperatures and measured at (a) $\VEC Q  = (1.7, 0, 0)$ and (b) $\VEC Q  = (0, 0, 1.9)$. 
    Solid lines represent fits as explained in the text.
    }
\end{figure}

Fig.~\ref{fig:rawdata}(a) shows the in-plane $\langle{\delta}{M_{b}}\rangle$ and out-of-plane $\langle{\delta}{M_{c}}\rangle$ fluctuations spectra at $\VEC Q = (1.8, 0, 0)$ in the maximum investigated temperature of 1.5$\times T_{C}$.
$\langle{\delta}{M_{b}}\rangle$ and $\langle{\delta}{M_{c}}\rangle$ where obtained by making the subtraction Eqs.~(6)-(4) and Eqs.~(5)-(4), respectively. 
We observe broad energy distributions centered at zero energy transfer (quasi-elastic excitations) that mark the existence of diffusive modes and are typical features of paramagnetic scattering~\cite{Moriya1985}.
Moreover, as can be seen at this temperature $\langle{\delta}{M_{b}}\rangle$ and $\langle{\delta}{M_{c}}\rangle$ are found to be identical pointing to isotropic fluctuations (Heisenberg spins).
In a previous study~\cite{Biniskos2017} similar behaviour was established for 1.036$\times T_{C}$ and therefore, one can assume that the dynamical spin susceptibility will remain isotropic in all the intermediate temperature range of $315\leq T \leq 457$\,K.
Fig.~\ref{fig:rawdata}(b) depicts subtracted energy spectra between two NSF channels ($\text{NSF}_{zz}$ and $\text{NSF}_{xx}$) acquired along the $(h00)$ direction at 1.2$\times T_{C}$.
As expected with increasing $q$ the intensity decreases while the signal broadens.
The spin fluctuations do not only change significantly with $q$, but are also strongly temperature dependent as illustrated in Fig.~\ref{fig:rawdata2}.
For a given scattering vector one observes a substantial diminution of the spectral weight with increasing temperature which is more prominent along the $(00l)$ direction. 
Following the evidence for isotropy in spin space, the data were collected only for $\langle{\delta}{M_{b}}\rangle$ and the outcome is described in the next section.

\begin{figure}[h]
    \includegraphics[width=\columnwidth]{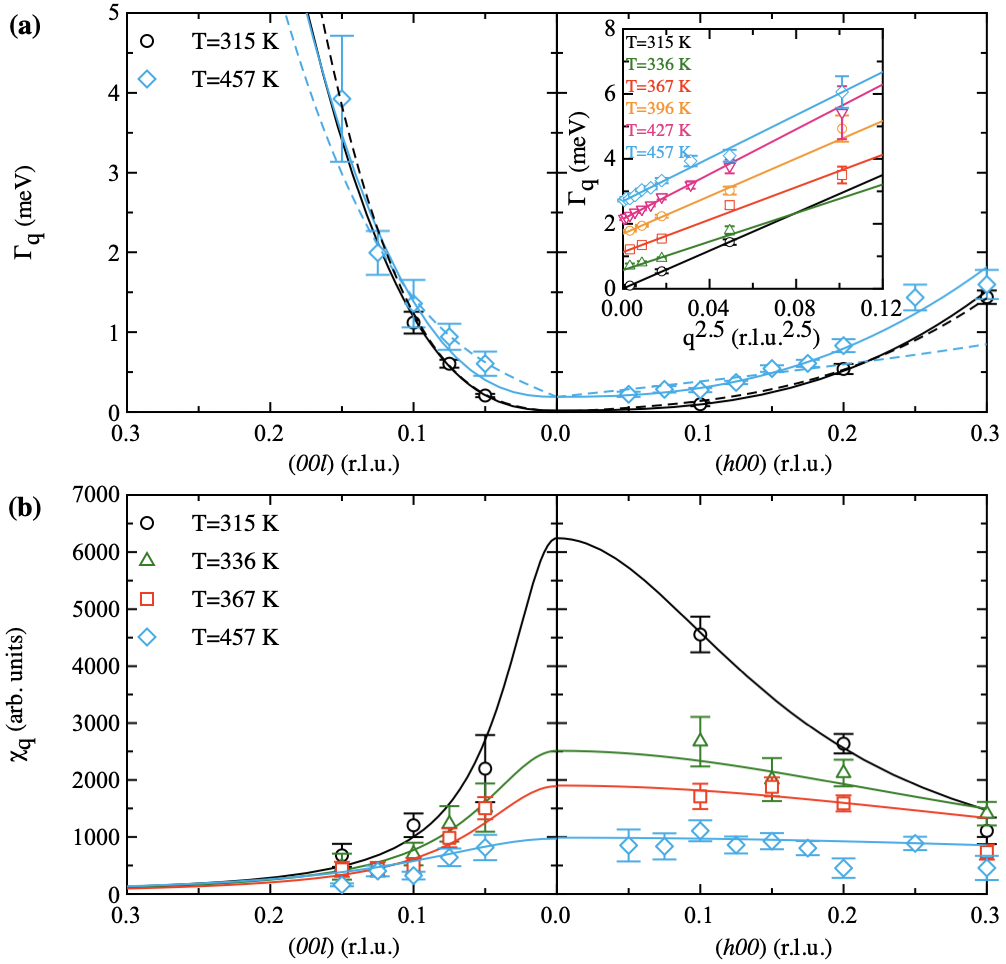}
    \caption{\label{fig:res1}
     (a) Linewidths $\Gamma_{q}$ and (b) $q$-dependent susceptibility $\chi_{q}$ of MnFe$_4$Si$_3$ obtained along the $(00l)$ and $(h00)$ directions at different temperatures. 
     For clarity only the results of $\Gamma_{q}$ obtained at the minimum and maximum investigated temperature are shown.
     The solid and dashed lines for $\Gamma_{q}$ correspond to fits with the models of Heisenberg and weak itinerant ferromagnetism, respectively, while for $\chi_{q}$ solid lines indicate fits with Lorentzian functions.
     The inset in (a) shows the linear behaviour for $\Gamma_{q}(T)$ vs $q^{2.5}$ for the $(h00)$ direction. For clarity, the data (in the inset only) are shifted by a constant value along the vertical axis. 
    }
\end{figure}

In order to analyze the obtained spectra we followed the same procedure which is given in detail in Ref.~\onlinecite{Biniskos2017}.
Briefly we mention that the subtracted intensities were convoluted with the 1D-instrument resolution in the energy direction and the PM scattering for the energy transfer between $-4\leq E \leq 4$\,meV can be described with the empirical double Lorentzian function~\cite{Endoh2006}:
\begin{eqnarray}
\label{SQE}
\begin{split}
\text{S}(\VEC Q , E) & = k_{B}T\frac{\chi_{0}}{1+ (q/\kappa)^2}\frac{\Gamma_{q}}{{E}^2 + \Gamma_{q}^2} \\
			& = k_{B}T\frac{\chi_{q}\Gamma_{q}}{{E}^2 + \Gamma_{q}^2},\\
\end{split}				
\end{eqnarray}
where $\chi_{0}$, $\kappa$, $\Gamma_{q}$, and $\chi_{q}$ are the static susceptibility, the inverse spin correlation length, the $q$-dependent energy linewidth (relaxation rate), and the $q$-dependent susceptibility, respectively.
The obtained values for $\Gamma_{q}$ and $\chi_{q}$ for the $(00l)$ and $(h00)$ directions at different temperatures are shown in Fig.~\ref{fig:res1}.

\begin{table*}[thb]
\begin{center}
\begin{ruledtabular}
\begin{tabular}{c c c c c c c} 
$T$ & $\Gamma_{0}$ & $A_{(h00)}$ & $A_{(00l)}$ & $\chi_{0}$ & $\kappa_{(h00)}$ & $\kappa_{(00l)}$ \\
(K) & (meV) & (meV{$\text{\AA}^{2.5}$}) & (meV{$\text{\AA}^{2.5}$}) & (arb. units) & ({$\text{\AA}^{-1}$}) & ({$\text{\AA}^{-1}$})  \\ [0.5ex] 
\hline
315 & 0.002(5) & 24.8(6) & 184(6) & 6200(1100) & 0.161(23) & 0.054(13) \\
336 & 0.050(21) & 20(3) & 155(18) & 2500(500) & 0.388(108) & 0.088(19) \\
367 & 0.087(17) & 24(3) & 179(14) & 1720(190) & 0.490(104) & 0.104(16) \\
396 & 0.119(25) & 28(2) & 164(17) & 1400(140) & 0.55(13) & 0.117(27) \\
427 & 0.135(22) & 31(2) & $-$ & 1040(150) & 0.800(211) & $-$ \\
457 & 0.190(29)  & 32(3) & 153(19) & 987(111) & 0.807(223) & 0.16(4) \\
\end{tabular}
\end{ruledtabular}
\caption{\label{tab:results}
Microscopic parameters describing the spin fluctuation spectrum of ferromagnetic MnFe$_4$Si$_3$ at different temperatures in the paramagnetic state. 
Data were not measured for the $(00l)$ direction at $T = 427$\,K.}
\end{center}
\label{default}
\end{table*}

\begin{figure}[h!]
    \includegraphics[width=\columnwidth]{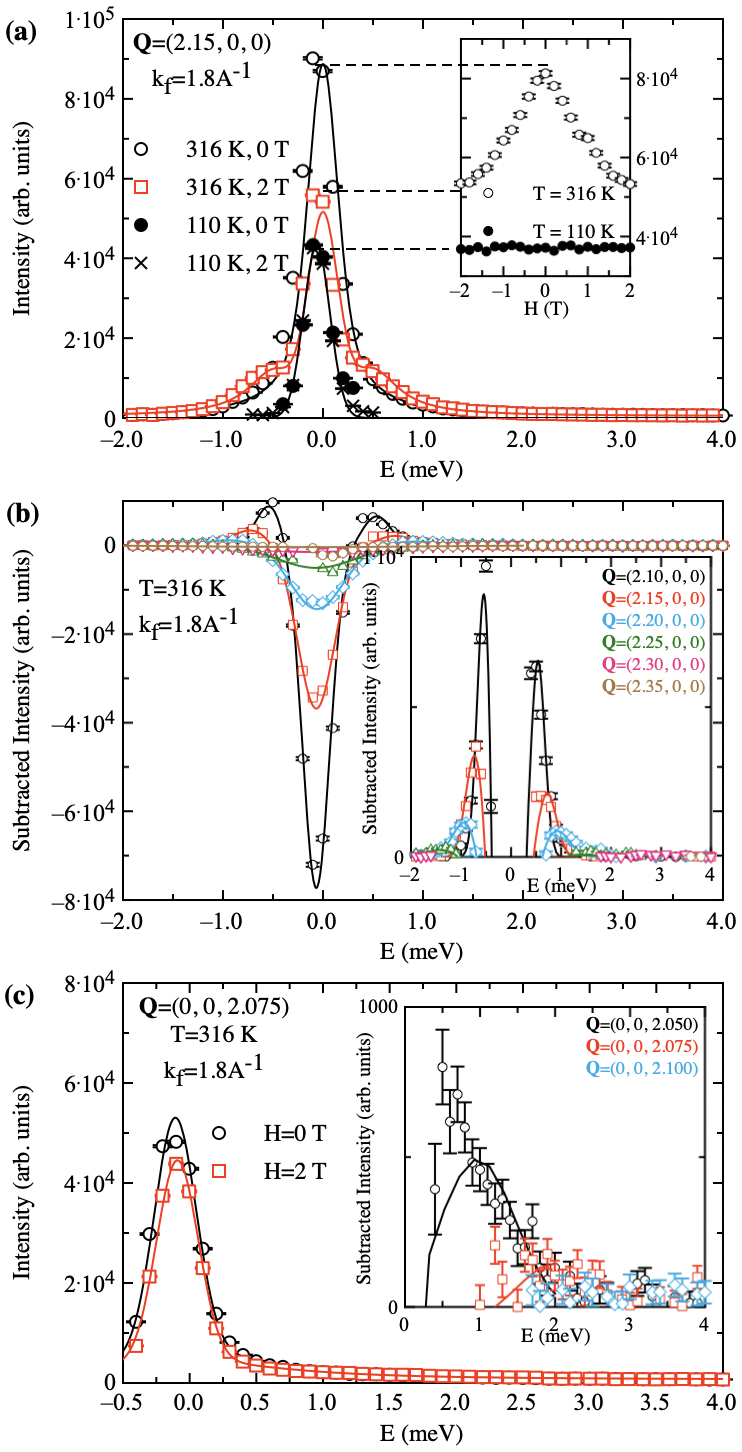}
    \caption{\label{fig:mag}
    Spin fluctuations spectra of MnFe$_4$Si$_3$ obtained at $H = 0$\,T and $H = 2$\,T  and measured at (a) $\VEC Q  = (2.15, 0, 0)$ and (c) $\VEC Q  = (0, 0, 2.075)$. 
    In panel (a), the signal obtained at 110\,K at 0\,T and 2\,T for $\VEC Q  = (2.15, 0, 0)$ indicates the incoherent background.
    The inset shows the field dependence of $\VEC Q  = (2.15, 0, 0)$ at zero energy transfer at 316\,K and 110\,K.
    (b) Subtracted intensities ($I(2\text{T})-I(0\text{T})$) at $T = 316$\,K at different $\VEC Q$ positions along the $(h00)$ direction. 
    The insets in panels (b) and (c) focus on the positive part of the subtracted intensities ($I(2\text{T})-I(0\text{T})$) at $T = 316$\,K for the $(h00)$ and $(00l)$ directions, respectively.
    Lines are guides for the eyes.
    }
\end{figure}

The $q$-dependent energy linewidth between the $(00l)$ and $(h00)$ high symmetry directions of the hexagonal system are strongly anisotropic (see Fig.~\ref{fig:res1}(a)) in line with the results obtained for the spin-wave spectrum measured below $T_C$~\cite{Biniskos2017}.
To describe the experimental data at $T = 1.5\times T_{C} = 457$\,K we use two different models~\cite{Endoh2006}: the expression for weak itinerant FM $\Gamma_{q}=\Gamma_{0} + A_{\text{wi}}q(1+(q/\kappa)^{2})$ and the empirical formula $\Gamma_{q} = \Gamma_{0} + Aq^{z}$, where $\Gamma_{0}$ refers to the $q=0$ intercept of $\Gamma_{q}$.
The data are better described by the latter formula and we obtain a finite value for $\Gamma_{0}$ and the exponent $z = 2.44(19)$.
The exponent is close to $z = 2.5$ which describes the relaxation rate of the magnetic fluctuations for Heisenberg ferromagnets~\cite{Endoh2006}.
In a previous study~\cite{Biniskos2017}, the experimental data for the relaxation rates at $T = 1.036\times T_{C} = 316$\,K could be well described by both models, the model for Heisenberg FM, as well as, the model for weak itinerant FM.
However, in the present work we demonstrate that at $T = 1.5\times T_{C}$ only the former model is suitable for $\Gamma_{q}$ (see solid lines in Fig.~\ref{fig:res1}(a)).
Consequently the model $\Gamma_{q}(T) = \Gamma_{0}(T) + A(T)q^{2.5}$ was used to fit the data in the present work for $T \leq 457$\,K.
A zoom of the low $q$ data obtained along $(h00)$ is shown in the inset of Fig.~\ref{fig:res1}(a) where $\Gamma_q$ is plotted as a function of $q^{2.5}$.

Fig.~\ref{fig:res1}(b) shows the temperature dependence of the $q$-dependent susceptibility $\chi_{q}$.
For all the investigated temperatures $\chi_{q}$ decreases faster along the $(00l)$ direction compared to $(h00)$ indicating a shorter inverse correlation length. 
We used a Lorentzian function for $\chi_{q}$, $\chi_{q}^{-1}$=$\chi_{0}^{-1}$(1+ $(q/\kappa)^2$) (see Eq.~\eqref{SQE}), in order to extract the values for the static susceptibility $\chi_{0}$ and the inverse correlation lengths $\kappa$.
All the experimentally obtained results of $\Gamma_{0}$, $A$, $\chi_{0}$ and $\kappa$ for each temperature are summarized in Table~\ref{tab:results}.
While $A$ depends substantially on direction in line with $\Gamma_q$, it is weakly affected by temperature. 
All other quantities in Table~\ref{tab:results} change significantly with temperature.
The $q=0$ susceptibility, $\chi_{0}$, decreases strongly with increasing temperature pointing to the criticality of spin fluctuations near $T_C$. 
However, its precise temperature dependence (not shown) does not follow a $1/T$-like behaviour in relation to the fact that the Curie-Weiss law was found to describe the bulk susceptibility data only for $T > 550$\,K~\cite{Hering2015}. 
$\Gamma_0$ follows an inverse temperature variation compared to $\chi_0$. 
Following the behaviour of  $\chi_q$, the inverse spin correlation lengths $\kappa$ are strongly temperature and directional dependent. 
The quantities $\kappa$, $A$ and $\Gamma_0$ that quantify the extend of spin fluctuation in ($\textbf{q}$, $E$)-space are viewed in detail in the Discussion section.

\subsection{Field dependence of paramagnetic scattering at 1.036$\times T_{C}$}

In a previous study, it was shown that a magnetic field of 2\,T suppresses the elastic contribution measured at low $Q$ close to $T_C$ and integrated in the energy range -0.1 to 0.1\,meV (see Fig.~6 in Ref.~\onlinecite{Biniskos2017}).
A similar measurement is shown for a higher $Q$ position in the inset of Fig.~\ref{fig:mag}(a) through a field dependence of the intensity for $\VEC Q  = (2.15, 0, 0)$ for zero energy transfer at 316\,K and 110\,K (with the experimental configuration presented above). 
The intensity at 316\,K decreases with magnetic field, while the measurement at 110\,K is field independent and corresponds to the incoherent background (note that this contribution is cancelled by subtraction in the polarized data shown in the previous subsection).
At $|H|=2$\,T, the remaining magnetic signal is above this background, however, a considerable amount ($\approx 2/3$) of magnetic elastic signal is suppressed. 
The amount of suppression is found to be $\VEC q$-dependent.

In order to further understand the field evolution of the magnetic excitation spectrum, the energy dependence of the fluctuations was comparatively studied at zero and finite magnetic field up to an energy transfer  of 4\,meV. 
Spectra along the $(h00)$ and $(00l)$ directions were collected at 316\,K at $H = 0$\,T and $H = 2$\,T. 
Representative measurements are shown in Fig.~\ref{fig:mag}(a) and Fig.~\ref{fig:mag}(c), for $\VEC Q  = (2.15, 0, 0)$ and $\VEC Q  = (0, 0, 2.075)$, respectively. 
Consistently with the polarized data (see Fig.~\ref{fig:rawdata} and Fig.~\ref{fig:rawdata2}), a signal centered at $E = 0$\,meV is observed for both directions at 0\,T. 
Applying a magnetic field of 2\,T results in a reduction of this PM scattering consistently with the field scan presented in the inset of Fig.~\ref{fig:mag}(a) as indicated by the dashed lines (for $\VEC Q  = (2.15, 0, 0)$, the incoherent background signal obtained at 110\,K for 0\,T and 2\,T is also shown). 
The $\VEC Q$ dependence of the subtracted intensities, $I(2\text{T})-I(0\text{T})$, is shown in Fig.~\ref{fig:mag}(b) for the $(h00)$ direction.
In the quasi-elastic regime (low energy transfers), it is negative and decreases in absolute value when $q$ increases. 
At finite energies and small wave-vectors, this difference is positive and the spectra show a local maximum at finite energy transfers which position disperses with respect to the wave-vector. 
This is better highlighted in the insets of Fig.~\ref{fig:mag}(b) and Fig.~\ref{fig:mag}(c) for the $(h00)$ and $(00l)$ directions, respectively. This peak at finite energy corresponds to spin waves associated with the fact that the paramagnetic system re-enters the ferromagnetic phase under finite magnetic field. 
Such phenomena are rarely reported in neutron scattering studies, examples being the FM Gd~\cite{Cable_1989} and EuS~\cite{Rebelsky1990}.

\section{Discussion}

\subsection{Nature of spin fluctuations}

\begin{figure}[h]
    \includegraphics[width=\columnwidth]{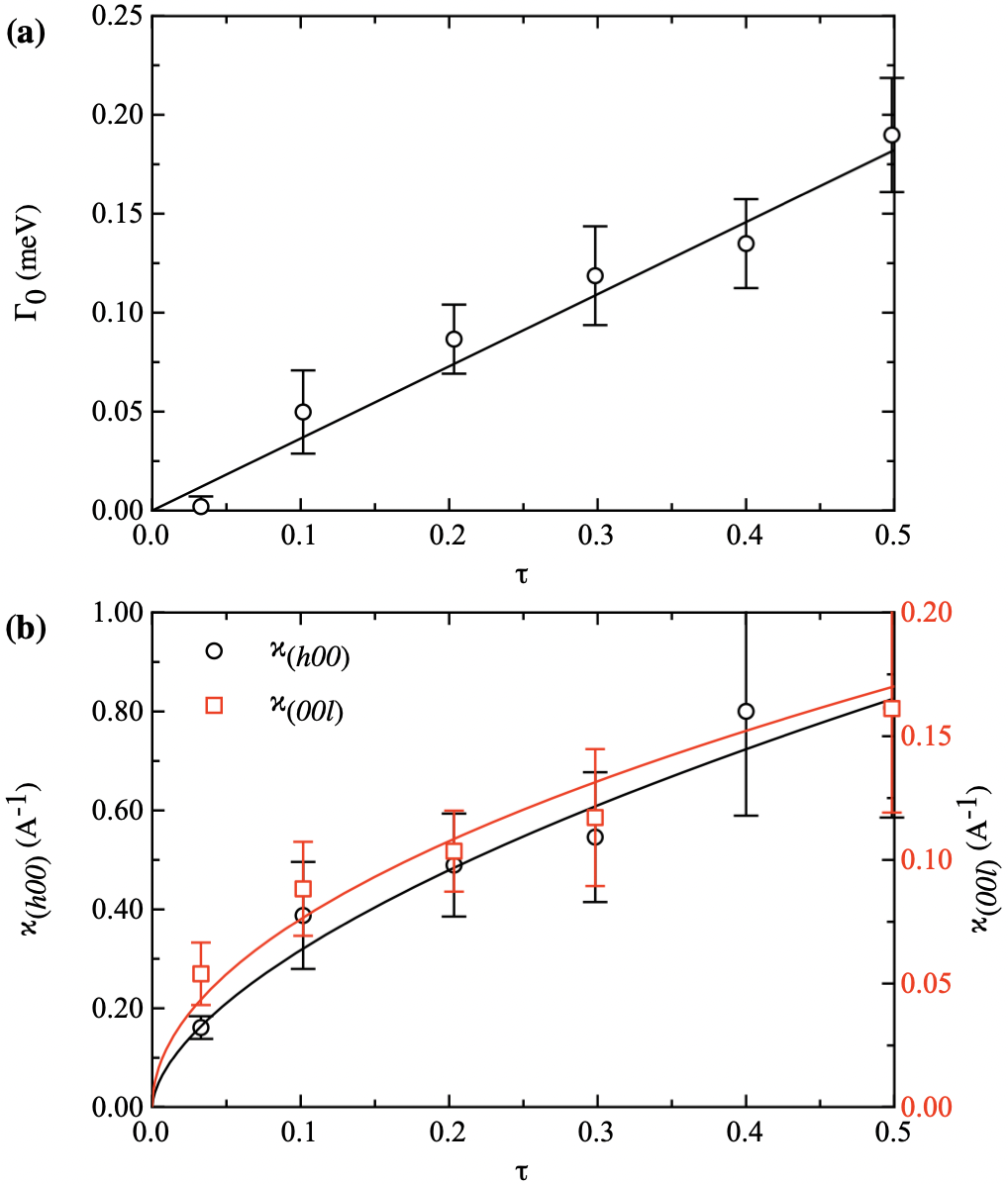}
    \caption{\label{fig:res2}
     Temperature dependence of (a) $\Gamma_{0}$ and (b) the inverse spin correlation lengths $\kappa$ along $(h00)$ (left axis scale) and $(00l)$ (right axis scale) of MnFe$_4$Si$_3$.
     $\tau$ is the reduced temperature. 
     Solid lines represent fits as explained in the text.
    }
\end{figure}

Firstly, by analyzing the results for $\kappa$ we acquire information of the typical length scales of the system. 
For the inverse of the spin correlation lengths one can assume that they follow the critical law $\kappa = \kappa_{0} \tau^\nu$~\cite{Endoh2006}, where $\tau = (T/T_{C}-1)$ is the reduced temperature.
Such a fit yields $\kappa_{0}^{(h00)} = 1.25(24)$\,$\text{\AA}^{-1}$, $\kappa_{0}^{(00l)} = 0.24(2)$\,$\text{\AA}^{-1}$, and $\nu = 0.56(8)$ (see Fig.~\ref{fig:res2}(b)).
The obtained values for $\kappa_{0}$ are close to the ones previously extracted from the spin-wave stiffness $D$, using the relation $\kappa_0^2=3k_BT_C/(S+1)D$~\cite{Biniskos2017,Biniskos2018}, while $\nu$ approaches the value of 0.5 which corresponds to the exponent in the mean-field approximation.
For MnFe$_4$Si$_3$ when considering the lattice parameters~\cite{Hering2015}, it becomes evident that the interatomic distances are comparable to $\kappa_{0}^{-1}$, which points to a localized magnetic system.

Secondly we analyze the results for the relaxation rates in view of extracting the characteristic energy scales.
As discussed in the experimental results section, the data can be well described  in the whole investigated temperature range by the formula  $\Gamma_{q}(T) = \Gamma_{0}(T) + A_{q}(T)q^{2.5}$. 
It is found that the $q=0$ relaxation rate, $\Gamma_0(T)$, increases linearly with temperature and vanishes at $T_{C}$, following $\Gamma_{0}(\tau) = 0.36(3)\tau$ (see Fig.~\ref{fig:res2}(a)). 
Given the relation between $\kappa$ and $\tau$, we consequently obtain $\Gamma_{0} \propto \kappa^2$. 
This specific law is the signature of pseudo-dipolar interactions (interplay between crystal field and spin-orbit interactions) being responsible for the finite relaxation rate at $q=0$~\cite{Huber1983,MezeiF1984}. 
When $\Gamma_q$ is finite at $q=0$, the spin dynamics is so-called non conserved since the conservation of magnetization (i.e. the magnetization commutes with the Hamiltonian), that implies $\Gamma_{q=0}=0$, is necessarily broken. 
The coefficient $A$ is weakly temperature dependent, which makes it a pertinent parameter to quantify the spread of spin fluctuations in $E$-space. 
Since the exponents linking $\Gamma_{0}$ with $\kappa$, and $\Gamma_{q}$ with $q$, are different, the data cannot collapse on a single scaling function of the form $\Gamma_{q}(T)=Aq^zf(\kappa/q)$, with all the $T$ dependence being included in $\kappa$. 
It is recalled that when such scaling is valid, $f$ follows a non-monotonic R\'esibois-Piette function~\cite{Resibois1970} for the Heisenberg model, while in itinerant magnets, $f$ is monotonic with $f(\kappa/q)=(1+(\kappa/q)^2)$~\cite{Endoh2006}. 
At $T_C$, where $\kappa=0$, $\Gamma_{q}=Aq^z$ with $z=2.5$ for Heisenberg systems and $z=3$ for itinerant magnets~\cite{Moriya1985,Endoh2006}. 
The behaviour at $T_C$ is persistent even in the absence of scaling since $\Gamma_{0}(T_C)=0$. 
It was found experimentally that the exponent $z=2.5$ works fairly well even for the archetypal itinerant magnet Ni$_{3}$Al~\cite{Semadeni2000} and to a less extend for MnSi~\cite{Ishikawa1985} (note that MnSi is not a ferromagnet but an helimagnet with a very long period). 
Therefore, a quantitative comparison of the spin fluctuations of different isotropic ferromagnets can be made using an exponent $z=2.5$ to describe the $q$ dependence of the relaxation rate near $T_C$. 

\begin{table*}[thb]
\begin{center}
\begin{ruledtabular}
\begin{tabular}{c c c c c c c c c c} 
material & space group & $k_{B} T_{C}$ & $A_{0}$ & $\Gamma_{\text{ZB}}$ & $\Gamma_{\text{ZB}} / k_{B} T_{C}$ & $\kappa_{0}^+$ & lattice~constant & $\Delta T_{\text{ad}}/H$ \\
 & & (meV) & (meV{$\text{\AA}^{2.5}$}) & (meV) & & ($\text{\AA}^{-1}$) & ($\text{\AA}$)  & (K/T) \\
[0.5ex]
\hline 
Ni$_3$Al & $Pm\bar{3}m$ & 6.3 & 226(7)~\cite{comment2} & 390 & 61.9 & 0.049(19)~\cite{Semadeni2000} & 3.57 & -- \\
Ni & $Fm\bar{3}m$ & 54.6 & 367(18)~\cite{Boni1993} & 1075 & 19.7 & 0.81(4)~\cite{Boni1991} & 3.54 & 0.68~\cite{Gschneidner2000} \\
CoS$_2$ & $Pa\bar{3}$ & 10.4 & 86(2)~\cite{Hiraka1997} & 82.9 & 7.97 & 0.232~\cite{Hiraka1997} & 5.52 & 1.2~\cite{Wada2006,Mishra2019} \\
MnSi & $P2_{1}3$ & 2.6 & 19.6~\cite{Ishikawa1985} & 18.4 & 7.08 & 0.18~\cite{Ishikawa1985} & 4.56  & 1.3~\cite{Arora2007} \\
Co & $Fm\bar{3}m$ & 119.5 & 300(30)~\cite{Glinka1977} & 837 & 7.01 & 0.98(4)~\cite{Schinz1998} & 3.61 & 1.4~\cite{Gschneidner2000} \\
$\alpha$-Fe & $Im\bar{3}m$ & 89.8 & 135(5)~\cite{Mezei1982} & 402 & 4.48 & 0.82(3)~\cite{Collins1969} & 2.87 & 2.1~\cite{Gschneidner2000} \\
Pd$_2$MnSn & $Fm\bar{3}m$ & 16.4 & 60~\cite{Kohgi1986} & 40.3 & 2.46 & 0.22~\cite{Kohgi1986} & 6.38 & -- \\
EuO & $Fm\bar{3}m$ & 5.9 & 8.3(7)~\cite{Boni1986} & 9.57 & 1.62 & 0.64(4)~\cite{Boni1986} & 5.14 & 1.6~\cite{Kyunghan2005} \\
EuS & $Fm\bar{3}m$ & 1.4 & 2.1(3)~\cite{Boni1987} & 1.73 & 1.24 & 0.55(3)~\cite{Schinz1998} & 5.88 & 2.9~\cite{Li2014} \\
\hline
MnFe$_4$Si$_3$ & $P6_{3}/mcm$ & 26.3 & 24.8(6) & 5.14 & 0.20 & 1.25(24) & 6.81 ($a$-axis)& 0.59~\cite{Maraytta2019} \\
 & & & 184(6) & 66.2 & 2.52 & 0.24(2) & 4.73 ($c$-axis)& \\
\end{tabular}
\end{ruledtabular}
\caption{\label{tab:results2}
Microscopic parameters describing the spin fluctuations for various ferromagnets.
$\Gamma_{\text{ZB}}$ is the relaxation rate at the Brillouin zone boundary in the [111] direction for Ni, CoS$_2$, Co, Pd$_2$MnSn, EuO, and EuS, and in [110] for Ni$_3$Al, MnSi, and $\alpha$-Fe.
$\kappa_{0}^+$ refers to the inverse spin correlation length determined from measurements above $T_{C}$ (different values of $\nu$ are reported between the mean-field $\nu=0.5$ and the Heisenberg critical value $\nu=0.7$ depending on the systems and the temperature range studied). 
For a given material, the parameters reported in literature can be spread out due to different experimental conditions or different data analysis methods, and the references of the mentioned study are given for each value of the Table.
$\Delta T_{\text{ad}}/H$ is the estimated rate of change of the adiabatic temperature per Tesla at $T_{C}$ using the relevant references.
The results for MnFe$_4$Si$_3$ are obtained from this work and values are given for the [100] and [001] directions.
The $A_{0}$ values for MnFe$_4$Si$_3$ are given at 315\,K.
The symbol ``--'' indicates not reported values to our knowledge. 
The lines of the Table for the cubic ferromagnets are sorted by decreasing $\Gamma_{\text{ZB}}/k_BT_C$.
}
\end{center}
\label{default}
\end{table*}

For comparing the results of this study for MnFe$_4$Si$_3$ with the existing data for various isotropic cubic ferromagnets we adopt the approach of Ref.~\cite{Lynn1984}. 
From the results gathered in literature, we calculate the characteristic energies of each FM at the Brillouin zone boundary $\Gamma_{\text{ZB}}(T_C)=Aq_{\text{ZB}}^{2.5}$ (for MnFe$_4$Si$_3$ the given values are $\Gamma_{\text{ZB}}(315\,\text{K})$).
The results are summarized in Table~\ref{tab:results2} and the lines are arranged by decreasing $\Gamma_{\text{ZB}}/k_BT_C$, i.e. from more itinerant to more localized systems. 
It is known that for localized Heisenberg insulators (EuO and EuS) $\Gamma_{\text{ZB}} \approx k_{B} T_{C}$, while for metallic itinerant magnets $\Gamma_{\text{ZB}} >> k_{B} T_{C}$~\cite{Endoh2006}. 
From Table~\ref{tab:results2},  one concludes that MnFe$_4$Si$_3$ can be categorized as a localized FM system.  
It should be noted that the $q^{2.5}$ dependence of the relaxation rate is surprisingly found to be valid in the wave-vector range $\kappa \sim q$ for $(h00)$ and $(00l)$, as well as, in $\kappa >> q$ for $(h00)$, while it is expected to hold in the critical regime $q >> \kappa$. 
This latter regime cannot be reached in our experiment for different reasons depending on the direction probed: along $(h00)$ high values of $\kappa$ are obtained already at 315\,K, and along $(00l)$ the intensity drops too fast as a function of $q$. 
The hydrodynamic regime, corresponding to $\Gamma_{q} \propto q^{2}$ and expected for $\kappa >> q$, is hidden by the residual relaxation rate $\Gamma_{0} \propto \kappa^2$ due to the large $\kappa$~\cite{MezeiF1984}. 
To summarize this section, the extension of spin fluctuations both in $\textbf{q}$ and $E$ space, quantified by $\kappa_{0}$ and $A$, respectively, points towards localized magnetism for MnFe$_4$Si$_3$. 

\subsection{Comparison with other magnetocaloric compounds}

Within intermetallic compounds containing Mn and/or Fe, (Mn,Fe)$_{2}$(P,Si) and LaFe$_{13-x}$Si$_{x}$ stand out for their remarkable MCE characteristics~\cite{Bruck2024}. 
Both systems are considered as itinerant magnetic compounds although in (Mn,Fe)$_{2}$(P,Si), a scenario sometimes put forwards is the one of mixed magnetism, where local and itinerant moments coexist~\cite{Dung}. 
Sizeable spin fluctuations were evidenced by INS for both (Mn,Fe)$_{2}$(P,Si)~\cite{Miao_2016} and  LaFe$_{13-x}$Si$_{x}$~\cite{Faske_2020,Zhang_2021,Morrison2024}. 
These studies did not take the canonical approach followed in the present work and hence, did not provide microscopic parameters like the bare inverse spin correlation length $\kappa_0$ or the $A$ coefficient of the power law linking the relaxation rate to the wave-vector. 
In contrast to these itinerant magnets, we have shown by INS that the MnFe$_4$Si$_3$ system is clearly characterized by local magnetism above room temperature despite the occurrence of several magnetic sites that could favour mixed magnetism (WP 4$d$ non-magnetic site with Fe occupancy and WP 6$g$ magnetic site with mixed Fe/Mn occupancy~\cite{Hering2015}). 
It would be interesting to investigate the paramagnetic fluctuations in other promising FM compounds containing Fe or Mn which could be developed for magnetic refrigeration applications and spin dynamics investigations below $T_{C}$ have been already carried out, e.g., Fe$_2$P~\cite{Komura1980}, Mn$_5$Ge$_3$~\cite{dosSantosDias2023}, MnBi~\cite{Williams2016}, MnSb~\cite{Radhakrishna1996,Taylor224418}, and MnP~\cite{Yano2013}, in order to reveal their nature of magnetism.

\subsection{Spin fluctuations and magnetocaloric effect}

The magnetic excitation spectrum measured by neutron scattering above $T_C$ reflects the importance of thermal fluctuations to which the entropy is also related.
A quantitative connection can be achieved between the magnetic entropy and the spin fluctuation spectrum in the limit $T \rightarrow 0$\,K~\cite{Brinkman1968,Lonzarich1986}, which is very useful to relate the spin dynamics with the Sommerfeld coefficient obtained from specific heat measurements as demonstrated in correlated electron systems ~\cite{Hayden2000,Murani_2004,Lester2021}. 
However, such common approximation is not valid at high temperatures due to interactions between individual spin fluctuations.
We develop in Appendix B an alternate approach to relate the entropy $\mathcal{S}$ to the scattering function $S_{i}(\VEC Q, E)$ based on the neutron sum rule.
The obtained magnetic field derivative of the entropy $\mathcal{S}$ is:
\begin{equation}
\label{PMvsH}
\left(\frac{\partial \mathcal{S}}{\partial H}\right)_T = - \frac{1}{2MV^2} \frac{\partial}{\partial T} \sum_{i,\VEC Q}\int \limits_{-\infty}^\infty S_{i}(\VEC Q, E)dE ,
\end{equation}
where $i$=$(\parallel, \perp)$ corresponds to the response parallel or perpendicular to the applied magnetic field. 
This distinction arises because, the magnetic field necessarily breaks the isotropy of spin fluctuations~\cite{Lazuta}.  
In Fig.~\ref{fig:mag}, the data obtained at 2\,T are the sum of these contributions and a difficulty arises from the fact that they may have different characteristic linewidths with different temperature variations. 
This separation would be an enormous experimental task in terms of neutron beam time; basically the $H=0$ approach presented above should be repeated with polarized neutrons under field,  for the two channels $i=(\parallel, \perp)$.

In addition, as shown in the experimental section, two effects occur under field: (i) the low energy diffuse scattering (quasi-elastic signal) is suppressed at 2\,T and (ii) spin waves are induced at higher energy.
Analysing all these aspects in a quantitative way with the limited set of available data is beyond the scope of this paper.
Eq.~\eqref{PMvsH} indicates that the strong modification of the excitation spectrum under magnetic field is correlated with the change of magnetic entropy of the system and the magnetocaloric effect. 
Staying on qualitative grounds, our data show that the quasi-elastic signal is strongly affected by a field of 2\,T and that the field induced excitations are constituted of a dispersive mode. 
It can be safely inferred that a higher field will suppress all quasi-elastic signal in the benefit of the field induced spin-wave mode (and the field induced order parameter). 
Globally this should translate in a decrease of the magnetic entropy of the system under magnetic field, as experimentally reported in single crystals of MnFe$_4$Si$_3$ with an entropy change of about $-2$\,J/kgK at a field change from 0 to 2\,T along the [100] direction at 316\,K~\cite{Hering2015}.

\section{Conclusions}

By a detailed INS study of the spin fluctuation spectrum of MnFe$_4$Si$_3$, we have shown that the nature of the magnetism of this material is localized above room temperature. 
This conclusion stems from both the characteristic extend in wave-vector and energy space of the fluctuation spectrum. 
In the studied temperature range, $315\leq T \leq 457$\,K, the inverse correlation lengths $\kappa$ follow a temperature dependence close to the mean field behaviour (exponent $\nu=0.5$), and the bare inverse correlation lengths $\kappa_0$ are comparable to the interatomic distances. 
The relaxation rate follows $\Gamma_{q}(T) = \Gamma_{0}(T) + A(T)q^{2.5}$ in an extended $\bf{q}$-range, where the $z=2.5$ exponent is characteristic of Heisenberg ferromagnets. 
The $q=0$ intercept, $\Gamma_0$, vanishes at $T_C$ and follows $\Gamma_{0} \propto \kappa^2$, which indicates that pseudo-dipolar interactions are responsible of the non-conserved spin dynamics. 
The magnetic field response evidences a suppression of the quasi-elastic signal at 2\,T and the appearance of a magnetic field induced spin-wave mode at finite energy. 
This strong modification of the fluctuation spectrum is associated with the reduction of the system's entropy under magnetic field. 
While a quantitative analysis cannot be achieved in the present study, a route to link magnetic entropy and spin fluctuations is proposed for further experimental investigations which could be combined with theoretical modelling.
Undertaking canonical spin fluctuation studies in more magnetic systems will provide important information about microscopic magnetic properties, which in turn could help define strategies for improving specific materials for magnetic refrigeration applications.  

\section{Acknowledgements}

N.B.\ acknowledges the support of JCNS through the Tasso Springer fellowship and the Czech Science Foundation GA\v CR under the Junior Star Grant No. 21-24965M (MaMBA).
The neutron data collected at the ILL are available at Refs.~\onlinecite{data_IN22,data_IN12,data_IN12b}.

\section{Appendix A: Adiabatic temperature change of Table~\ref{tab:results2}}

For completeness, the adiabatic temperature change $\Delta T_{\text{ad}}$ corresponding to the materials temperature variation for an adiabatic magnetic field change $\Delta H$ is given for the different compounds listed in Table~\ref{tab:results2}. 
The available measurements of $\Delta T_{\text{ad}}$ are obtained with different $\Delta H$ and we estimate the rate of change of $\Delta T_{\text{ad}}$ per Tesla. 
This allows a comparison between the different systems, however, it should be noted that $\Delta T_{\text{ad}} \not\propto \Delta H$, but generally decreases as the applied magnetic field increases~\cite{Gschneidner2000}. 
When considering the infinitesimal adiabatic temperature change:
\begin{equation}
\label{MCE}
dT(T, H)=-\left( \frac{T}{C(T,H)} \right)_{H} \left( \frac{\partial{M(T,H)}}{\partial{T}} \right)_{H}dH ,
\end{equation}
where $M$ is the magnetization and $C$ the specific heat at constant pressure, it is obvious that the Curie temperature (or equivalently the operating temperature near $T_C$) and the specific heat will be determinant (and counterbalancing) factors. 
Given the fact that the factors entering in Eq.~\eqref{MCE} are highly materials dependant, it is not pertinent to extract a trend between microscopic magnetism and MCE over the variety of compounds listed in Table~\ref{tab:results2}. 
Indeed MCE properties are usually reviewed on experimental grounds considering the materials or the different families of materials one by one~\cite{Franco2012,Bruck2024,Gschneidner2000}. 

\section{Appendix B: Magnetic neutron scattering sum rule, fluctuations and entropy}

The isothermal entropy change $\Delta\mathcal{S}$ can be obtained through an integration of the infinitesimal change:
\begin{equation}
\label{test1}
d\mathcal{S}(T,H)= \left( \frac{\partial{\mathcal{S}(T,H)}}{\partial{H}} \right)_{T}dH ,
\end{equation}
which is often evaluated from the magnetization measurements using the Maxwell’s relation:
\begin{equation}
\label{test2}
\left( \frac{\partial \mathcal{S}(T, H)}{\partial H} \right)_{T} = \left( \frac{\partial M(T, H)}{\partial T} \right)_{H} .
\end{equation}
For localized magnetic system, the neutron sum rule is: 
\begin{equation}
\label{test3}
s(s+1) = \left<m\right>^2 + \sum_{i,\VEC Q}\int \limits_{-\infty}^\infty S_{i}(\VEC Q, E)dE ,
\end{equation}
where $s$ is the effective spin of the ground state and $\left<m\right>$ the ferromagnetic moment per site ($M$=$\left<m\right>/ V$, $V$ being the volume of the unit cell). 
The derivative of Eq.~\eqref{test3} gives an alternate way to estimate the change of entropy with field:
\begin{equation}
\label{test4}
\begin{split}
2MV^2 \left(\frac{\partial M}{\partial T}\right)_H & = 2MV^2 \left(\frac{\partial \mathcal{S}}{\partial H}\right)_T \\ 
& = - \frac{\partial}{\partial T} \sum_{i,\VEC Q}\int \limits_{-\infty}^\infty S_{i}(\VEC Q, E)dE . \\
\end{split}	
\end{equation}
This relation between magnetization, entropy and spin fluctuations is only valid for local magnetism since it derives from the sum rule in Eq.~\eqref{test3}. 
Interestingly, it shows that the change of entropy with field is related to the ratio of the temperature derivative of the wave-vector and energy integrated scattering function to the magnetization and, hence, it shows clearly the role of spin fluctuations that is implicit in Eq.~\eqref{test2}. 
It is indeed explicit from Eq.~\eqref{test4} that the temperature derivative of the magnetization is controlled by the magnetic excitation spectrum.
To our knowledge this relation was not considered so far in literature.

\bibliography{MnFe4Si3}

\end{document}